\documentclass[aps,prl,twocolumn,groupedaddress]{revtex4-1}
\usepackage{url}
\usepackage{latexsym}
\usepackage{amstext,amssymb,amsfonts,amsmath}
\usepackage{graphicx}
\usepackage{subfig}
\usepackage{epstopdf}
\usepackage{color}

\begin{document}

\makeatletter
\def\@biblabel#1{[#1]}
\makeatother

\markboth{Kosmas Kosmidis, Moritz Beber, Marc-Thorsten H\"{u}tt }{Network heterogeneity and node capacity lead to heterogeneous scaling of fluctuations in random walks on graphs}

\title{Network heterogeneity and node capacity lead to heterogeneous scaling of fluctuations in random walks on graphs
}
\author{Kosmas Kosmidis}
\address{School of Science and Engineering, Jacobs University Bremen\\
 28759 Bremen, Germany\\
k.kosmidis@@jacobs-university.de}

\author{Moritz Beber}
\address{School of Science and Engineering, Jacobs University Bremen\\ 28759 Bremen, Germany\\
m.beber@jacobs-university.de}

\author{Marc-Thorsten H\"{u}tt}
\address{School of Science and Engineering, Jacobs University Bremen\\
 28759 Bremen, Germany\\
 m.huett@jacobs-university.de}

\begin{abstract}
Random walks are one of the best investigated dynamical processes on graphs. A particularly fascinating phenomenon is the scaling relationship of fluctuations $\sigma  $ with the average flux $\langle f \rangle $. Here we analyze how network topology and nodes with finite capacity lead to deviations from a simple scaling law $\sigma \sim \langle f \rangle ^\alpha$. Sources of randomness are the random walk itself (internal noise) and the fluctuation of the number of walkers (external noise).  We obtained exact results for the extreme case of a star network which are indicative
of the behavior of large scale systems with a broad degree distribution.The latter are subsequently studied using Monte Carlo simulations. We find that the network heterogeneity amplifies the effects of external noise. By computing the `effective' scaling of each node we show that multiple scaling relationships can coexist in a graph with a heterogeneous degree distribution at an intermediate level of external noise. Finally, we analyze the effect of a finite capacity of nodes for random walkers and find that this also can lead to a heterogeneous scaling of fluctuations.

\end{abstract}


\maketitle

\section{Introduction}

Complex networks \cite{albert2002statistical,arenas2008synchronization,barabasi2002linked,boccaletti2006complex,bornholdt2003handbook,newman2010} consist of a fascinating research topic which has, during the last decade, revolutionized our understanding of dynamically interacting systems. 
Several complex computational and biological networks are, in fact, transport networks meaning that network edges serve as channels of flux towards selected nodes. The Internet, for example, serves as an information transport network with the edges transferring information flux, measured in bytes per second, from node to node. Thus, transport phenomena on complex networks comprise a characteristic research subfield which has attracted many researchers as, besides its theoretical interest, it has also important engineering applications \cite{bogachev2009occurrence,Chen20123336,tadic2007transport}.

For many complex, network-like systems the flow of material or information is an important functional feature \cite{boccaletti2006complex,Helbing:2006,Helbing:2009}. Examples include the flow of carbon atoms (in the form of diverse chemical compounds) through metabolic networks \cite{Matthaus:2008p4186,sonnenschein2012}, the flow of information through the network of internet routers \cite{Smith:2011ge,Vazquez:2002bo}, traffic flow in streets or roads \cite{Becker:2011,Holme:2003vz,Vespignani:2009bf} and via train connections \cite{Fretter:2010p1726}, the flow of material through machine networks in industrial production \cite{Becker:2012,Becker:2011} and many more. 

As a consequence, predicting or understanding the pattern of fluxes (i.e. the material/information transfer through all nodes in the network) in terms of the network topology is a principal goal in the investigation of dynamics on graphs. 

It has been observed, for example, that the distribution of metabolic fluxes in the bacterium Escherichia coli follows a power law, similar to the degree distribution of the underlying metabolic networks \cite{Almaas:2004:Nature:14985762}. Maps of random walks reveal the community structure in complex networks \cite{RosvalBergstrom2008,SchulmanGaveau2005}. In scale-free graphs, for example, excitations can self-organize into wave-like patterns around hubs \cite{Mueller-Linow:2008}. A network derived from passenger flow can serve as a foundation for predicting the spread of epidemic diseases \cite{Brockmann:2013bq}. Clearly, many differences exist among the above systems e.g., on the level of conservation laws, the typical signal-to-noise ratio, system size and the relevant architectural properties of the corresponding networks. 

On the theoretical level, over the last decade substantial progress has been made in understanding such flow on graphs. Flux-balance analysis, a method for predicting the steady-state distribution of metabolic fluxes for a given metabolic network from nutrient availability and an objective function (e.g., maximizing growth rate) is a highly successful tool for distinguishing between lethal and viable mutants in simple organisms \cite{Edwards:2001p763,price2004gsm}. This method has applications as far reaching as the metabolic states of human cells \cite{Duarte:2007p208}. Its' variants have been used to study which network features are enhanced during a simulated evolution of simple flow networks, when requiring robustness against link or node removal \cite{kaluza2008scn} or as a function of task complexity \cite{Beber-et-al2012a}.

A strategy for investigating, how network topology affects the flux pattern, is to explore simple dynamics on graphs, serving as benchmarks for these investigations. Random walks are an important class of such benchmark models which are proven very useful for disentangling the universal network effects from those effects specific to individual application systems \cite{duch2006scaling,Holme:2003vz,Simonsen:2004hb}. We would like to stress that for random walks on graphs, in addition to numerical simulations, also a substantial number of mathematical results exists (see, e.g., the review by Lovasz \cite{lovasz1993random}).   

 In a relatively recent article \cite{de2004fluctuations} several real transport networks were studied including the Internet, microprocessor networks, river streamflow networks and highway networks among others.
 The authors were interested in  the relationship between the average number of packets $\langle f_i \rangle$ arriving at a node $i$ during a certain time interval i.e. the average node flux, and the standard deviation $\sigma_i$ of this quantity.
 They find that $\langle f_i \rangle$ has a broad distribution and that $\sigma_i$ scales as a power law with the average flux, i.e.
\begin{equation}
\sigma \sim \langle f \rangle ^\alpha
\label{eq1}
\end{equation}  

It was suggested \cite{de2004fluctuations} that physical systems are divided in roughly two classes, those with $\alpha=1/2$ and those with $\alpha=1$. The  power law behavior with $\alpha=1/2$ is attributed to ``internal'' fluctuations while the $\alpha=1$ is related to strong ``external'' noise. Internal noise is caused by the fact that the system itself consists 
of discrete units and it is inherent in the very mechanism by which the system evolves. External noise denotes fluctuations created in a system 
by the application of a random force, whose stochastic properties are supposed 
to be known. 
Subsequent studies \cite{duch2006scaling,eisler2008fluctuation,prignano2012exploring} refined this view in the following way \cite{meloni2008scaling}: The relation between the dispersion and the nodes fluxes can be separated in two parts, one due to ``internal'' and the other due to  ``external'' noise.

\begin{equation}
  \sigma_i^2=\langle f_i \rangle + c (\langle f_i \rangle)^2
  \label{eq1b}
  \end{equation}
where the theoretically predicted value of the parameter $c$ is equal to zero in the absence of external noise.

An important question here is: When can one call data `represented by / obeying a power law '\cite{stumpf2012critical}? It is therefore instructive to look in more detail at the respective explicatory power of the two `models', the highly aggregated single scaling law parametrization given by Eq.\ref{eq1}  and the overlay of two scaling relationships as represented by Eq.\ref{eq1b}. In any case, the main idea that the scaling relation between the dispersion and the nodes fluxes can be used as a means to study the collective dynamical properties of a large network and the interplay between ``internal'' and ``external'' noise is rather appealing and has triggered a substantial amount of applied research \cite{almaas2006power,arenas2004community,barthelemy2010disentangling,hu2007effect,hutt2005interplay,kampf2012fluctuations,kishore2011extreme,xia2010optimal}. 

In this paper we are particularly interested in the effect of network heterogeneity on the dispersion-flux relation. We use scale-free networks with degree distribution $ P(k)\sim k^{-\gamma} $ for different $\gamma$ which allow us to model various ranges of heterogeneous structures since networks with $\gamma=2$ have a large number of very highly connected nodes while networks with $\gamma=3$ have much fewer nodes with considerably lower maximum degree. 
Such networks are markedly different from random networks or lattices where all nodes have statistically the same properties and ,hence, their topology is uniform and homogeneous.  
We use a model of multiple walkers performing fixed length random walks on network structures, similar to the model proposed in \cite{de2004fluctuations}. The number of walkers $W$ may be either fixed or a random variable uniformly distributed in the interval $[W-\Delta W,W+\Delta W ]$. We show that the model is exactly solvable for the star network and this solution provides valuable insights for the
dispersion-flux relation on scale-free networks with $2<\gamma<3$.  Based on the intuition provided by the results for the star network we performed Monte Carlo simulations of random walks on scale-free networks. We confirm that the random walk model leads to dispersion-flux power law scaling with $\alpha=1/2$ when $\Delta W=0$, to $\alpha=1$ when $\Delta W$ is large and we are able to observe intermediate exponents for a capable $\Delta W$ spectrum.
We also find, that an alternative analysis form (Eq. \ref{eq1b}) is a rather effective way of data analysis and we show that network heterogeneity enhances the transition to the $\alpha=1$ regime requiring lower $\Delta W$ for lower $\gamma$ exponents. Finally, we show that simple modifications of the random walk model with the inclusion of a node capacity or excluded volume interactions lead to regimes with non-power law dispersion-flux scaling.

\section{Methods}
\subsection{Random walk model}
In order to study how Eq.\ref{eq1} arises and its range of validity we use a dynamical model initially proposed in \cite{de2004fluctuations} and subsequently used in \cite{duch2006scaling,prignano2012exploring}. 
We study the diffusion of multiple walkers starting with a network of size $N$.
We use scale-free networks as the diffusion substrate in order to ensure a broad degree distribution.
In the simplest version of this model, we assume that a number of $w$ walkers are randomly placed on the nodes of the network. This number $w$ is the realization of a random variable chosen  with uniform probability from the range $[W-\Delta W, W+\Delta W ]$. Each node has a capacity $C$ to accept walkers, where $C$ is the number of walkers that can simultaneously be on the same node. In the case that $C$ is larger than the total number of walkers the problem is equivalent to the diffusion of $w$ independent walkers. When $C=1$ the problem is equivalent to diffusion with excluded volume interactions. Unless explicitly stated otherwise the results of our simulations are for unlimited capacity $C$. Each site has a counter, initially set to zero, which records the number of arrivals on it. We allow the walkers to perform $M$ steps each. Then we record the values of each site's counter which are our ``daily'' fluxes, we reset the counters to zero and start again for $D$ ``days''. We calculate the average flux and standard deviation for each node. Figure \ref{fig1}A shows Monte Carlo Simulation results on a scale-free network with $\gamma=3.0, k_{min}=2$ and size $N=10000$. We allowed $W=5000$ walkers to simultaneously perform 100 independent $M=1000$ step walks. We plot the `time series' of the number of the walker arrivals at the end of each $M$-step walk (i.e. 100 ``days'' in this case) at the counters of 2 sample nodes with degree $k=46, 30$ respectively. The figure also indicates the average flux $\langle f \rangle$ and standard deviation $\sigma$ for each of the 2 nodes. The main topic of this paper consists in the analysis of the fluctuations of such `time series'.    

\subsection{Exact results for the star network}
\begin{figure}
    \begin{center}
       \includegraphics[angle=0,width=8.5cm]{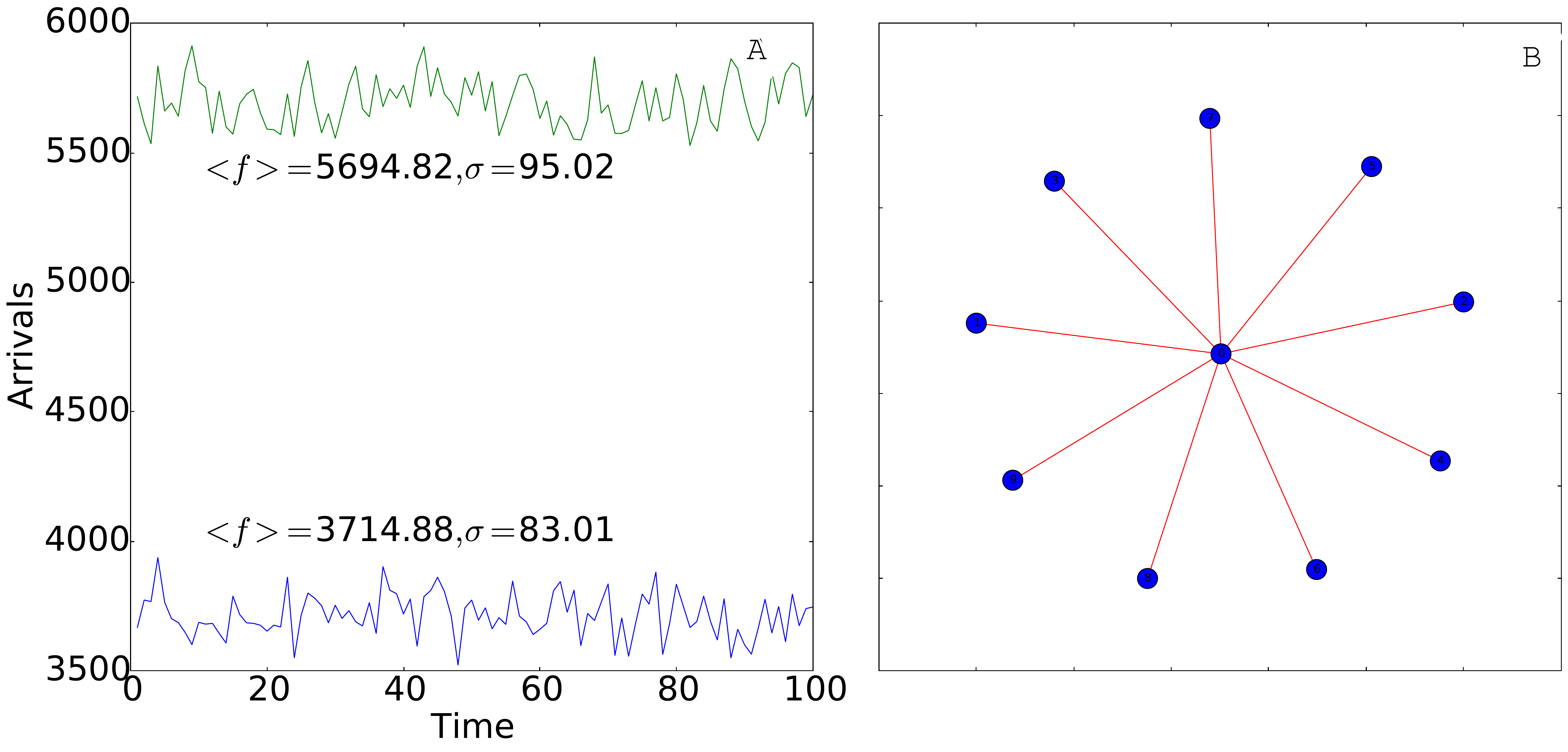}
    \end{center}
     \caption{(A) Monte Carlo Simulation results on a scale-free network with $\gamma=3.0, k_{min}=2$ and size $N=10000$. $W=5000$ walkers performed 100 $M=1000$ step walks. Lines are the `time series' of the walker arrivals at the counters of 2 sample nodes with degree $k=46, 30$ respectively. (B) A star network with $N=10$ nodes. Node 0 is the central hub with degree $k_0=9$. All other nodes have degree $k=1$.}
     \label{fig1}
  \end{figure}

Initially we study the walker dynamics on a simple construction such as the star network shown in Fig.\ref{fig1}A. This is a simple case of a bipartite graph with one central `hub' node (node 0) with degree $k_0=9$ and 9 other nodes with degree $k=1$ directly connected to node 0.
  
  The dynamical problem we have described is exactly solvable in the case of the star network. Without loss of generality we may
  assume that the number of steps $M$ is an even number. Let as initially consider the case where $\Delta W=0$ and $W=1$, i.e. a single walker performing $M$ steps on the star network.
  In such a case the flux of node 0 becomes deterministic because the node will be visited in every second step. Thus, if the walker is initially placed on any of the nodes 1 to 9, node 0 will be visited $M/2$ times (at steps $2,4,...,M/2$). If the walker is initially placed on node 0 then the node will again be visited $M/2$ times (at steps $1,3,...,M/2 -1$). Thus, $\langle f_0 \rangle=M/2$ and $\sigma_0=0$ where the averages are over different realizations of the $M$ step walk. The fact that the dynamics of the central node become deterministic ($\sigma_0=0$) allow us to calculate the probability that a peripheral node is visited $m$ times. The central node is accessed exactly $M/2$ times and each time the peripheral node $i$ will be visited with probability $p=1/k_0$ i.e. $p=1/9$ since node 0 has $k_0=9$. Thus, the probability $q(m)$ that a node $i$, ($i=1,2,...,9 $) is visited exactly $m$ times is given by a Binomial distribution
  \begin{equation}
	  q(m;M/2,p)={M/2 \choose m} p^m (1-p)^{M/2-m}
  \label{eq2}
  \end{equation} 
  
 The case of $W>1$ non-interacting walkers, when the network nodes have unlimited capacity $C$ can be viewed as equivalent to one walker performing
 $WM$ steps, leading to
 
 \begin{equation}
   q(m;WM/2,p)={WM/2 \choose m} p^m (1-p)^{WM/2-m}
   \label{eq3}
 \end{equation} 
 
 The mean number of visits on the peripheral nodes and the variance $\sigma^2$ are given from the well known formulas for the Binomial distribution leading to
 \begin{equation}
 f_i=\frac{WM}{2} p
 \label{eq4}
 \end{equation}
 
 \begin{equation}
  \sigma_i^2=\frac{WM}{2} p(1-p)
  \label{eq5}
  \end{equation}
    
 Thus, for the case of $\Delta W=0$ we recover the power law dispersion-flux scaling with exponent $\alpha=1/2$ as can readily be seen from eqs \ref{eq4}-\ref{eq5}.  
 
For discussing dynamics with ``external'' noise we study the case $\Delta W \neq 0$. Let us, for specificity, examine the case $\Delta W=1$. In this case, the number of walkers on our system is a random variable with discrete range $W-1,W,W+1$ and probability $1/3$ for each of these values.
 
 The resulting distribution for the flux on node $i$ is a mixture distribution i.e. the probability distribution of a random variable whose values can be interpreted as being derived from an underlying set of other random variables each with `weight' $w=1/3$ in this case.
 
 In case of a mixture of one-dimensional normal distributions with weights $w_i$, means $\mu_i$ and variances $\sigma_{i}^{2}$, the total mean and variance will be:
 
 \begin{equation}
 E[X] = \mu = \sum_{i = 1}^n w_i \mu_i 
 \label{eq6}
 \end{equation}
 
 \begin{equation}
 E[(X - \mu)^2] = \sigma^2 = \sum_{i=1}^n w_i((\mu_i - \mu)^{2} + \sigma_i^2)  
 \label{eq7}
 \end{equation}
 
where $ E[X]$ denotes the expected value of random variable $X$. For the central node we do not have to resolve to the above formula because $\sigma^2=0$ and we can calculate the resulting variance from the variation of walkers in a straightforward manner. This calculation yields identical results to those obtained with the use of the above eqs.\ref{eq4}, \ref{eq5}. 

In the following Table \ref{table1} we present results for the mean flux and dispersion for a set of $W=5$ random walkers performing $M=10$ step walks on the star network pictured in Fig.\ref{fig1}. Exact results were obtained from eqs. \ref{eq4}-\ref{eq7} and are in excellent agreement with Monte Carlo simulations of these walks on the star network.

  \begin{table}[h]
  \centering
  \begin{tabular}{|c|c|c|c|c|}
  \hline
   ~~ & \multicolumn{2}{|c|}{$\Delta W=0$}& \multicolumn{2}{c|}{$\Delta W=1$} \\
  \hline
  Node id& $\langle f \rangle $& $\sigma^{2}$& $\langle f \rangle,  $&$\sigma^2 $\\
  \hline
  node 0& 25 & 0 & 25 & 4.08 \\
  \hline
   nodes 1-9 & 2.77 & 1.57 & 2.77 & 1.63 \\
   \hline
   
  \end{tabular}
	\caption{Mean flux and dispersion for a set of $W=5$ random walkers performing $M=10$ step walks on the star network pictured in Fig.\ref{fig1}. Numbers are exact results using eqs. \ref{eq4}-\ref{eq7} and are in excellent agreement with Monte Carlo simulations of these walks on the star network.} 
   \label{table1}
   \end{table}

  These results give us immediately an indication of the effect of the ``external'' noise on the dynamics. First, we observe that the high degree node gets the majority of the flux. Moreover, we notice that the mean flux does not change with the introduction of a non zero $\Delta W$. Finally, we observe that the impact of $\Delta W$ on the variance is strongly dependent on the degree of the node. The low-degree nodes have a very mild increase of their variance (from 1.57 to 1.63 in our case-study) while the variance of the `hub' node jumps from 0 to 4.08.
  
  The exact results on the star network give us a hint on the origin of the $\alpha=1/2$ exponent, which is to be expected in processes where the observable random variable follows a Binomial (or Poisson) distribution as seen in our case study. We can also understand the origin of the $\alpha=1$ exponent from the following argument. Let $F$ denote the total flux on the network i.e. the sum of the flux on all nodes. When $\Delta W=0$, $F$ has a fixed value equal to the product $WM$, i.e. the total number of steps from all walkers on the network. When $\Delta W \neq 0 $, $F$ becomes a random variable. The total flux is distributed over the $N$ network nodes (indexed from 0 to $N-1$)
  \begin{equation}
  \langle F \rangle= \sum_{i=0}^{N-1} \langle f_i \rangle
  \label{eq8}
  \end{equation}
  We set $ B_i=\sum_{j \neq i} \langle f_j \rangle \Rightarrow B_i \leq \langle F \rangle$. We write $B_i=\alpha_i \cdot \langle F \rangle$ with $0 \leq \alpha_i \leq 1 $.
  Then,
  \begin{equation}
  \langle f_i \rangle =\alpha_i \langle F \rangle
  \label{eq9}
  \end{equation}
  For example, in our case-study for the `hub' of the star network $\alpha_0=1/2$ since $ \langle f_i \rangle=25$ and $ \langle F \rangle =50$. This holds for every $\Delta W$. For the rest of the nodes Eq.\ref{eq9} and Table \ref{table1} lead to, $\alpha_i \cdot 50=2.77 \rightarrow \alpha_i=0.055$. 
  When $\Delta W=0$, this reduces to $\langle f_i \rangle =\alpha_i F=\alpha_i MW$.  
  
  We know, however, that if $Y,X$ are random variables and $Y=aX$ where $a$ is a constant then the variances of the two are connected by $\langle \langle Y^2 \rangle \rangle =a^2 \langle \langle X^2 \rangle \rangle $, where we use the double bracket notation to denote the variance i.e. $\sigma_Y^2=\langle \langle Y^2 \rangle \rangle=\langle Y^2 \rangle-\langle Y \rangle ^2$.
  Thus, when $F$ is variated by introducing $\Delta W > 0$ we expect from Eq.\ref{eq9} 
  \begin{equation}
  \langle \langle f_i^2 \rangle \rangle=\alpha_i^2 \langle \langle F^2 \rangle \rangle
  \label{eq10}
  \end{equation}
  
  From eqs.\ref{eq9}-\ref{eq10} we obtain:

  \begin{eqnarray}
  \langle \langle f_i^2 \rangle \rangle&=&\left[ \frac{\langle f_i \rangle}{\langle F \rangle} \right] ^2 \langle \langle F^2 \rangle \rangle \Rightarrow \\
  \langle \langle f_i^2 \rangle \rangle &=& \frac{\langle \langle F^2 \rangle \rangle}{\langle F \rangle ^2} \langle f_i \rangle ^2  \Rightarrow \\
  \sigma_i&=& \sqrt{\frac{\langle \langle F^2 \rangle \rangle}{\langle F \rangle ^2}} \langle f_i \rangle  
  \label{eqa13}
  \end{eqnarray}
  
  leading to the desired scaling form. Eq.\ref{eqa13} can be readily verified for the star network using the results of Table \ref{table1}.

\section{Monte Carlo Simulations and Results}
The above results give us an indication of the way the two commonly observed scaling exponents appear and a hint for the mechanism of the transition from the one limiting case to the other. They are a helpful guide for the computational study of larger networks. We have seen that $\alpha=1/2$ scaling is to be expected when our observable quantity $Y$ follows a Binomial or Poisson distribution while $\alpha=1$ scaling arises in the case of random variables $Y,X$ having a multiplicative relation $Y=bX$ with constant $a$ since then $\langle Y \rangle =b \langle X \rangle $ and $\langle \langle Y^2 \rangle \rangle=b^2 \langle \langle X^2 \rangle \rangle $ which eliminating $b$ will lead to $\langle \langle Y^2 \rangle \rangle =\frac{\langle \langle X^2 \rangle \rangle}{\langle X \rangle^2} \langle Y \rangle^2$.

We have also seen that large-degree nodes are more sensitive when $\Delta W>0$  i.e. they are more sensitive to external noise. Hence, a plausible assumption on the influence of network structure on the dynamics is the following. Due to the network heterogeneity a broad flux distribution becomes observable since well-connected nodes get more flux than low-connected ones. The presence of external noise influences the nodes in a non-uniform way, with highly-connected node fluctuations enhanced much more, and thus leading to a change of the dispersion-flux relation.   

To verify this assumption for larger networks we have simulated diffusion on the largest connected component of scale-free networks with nodes $N=10^4$ and $\gamma=2.0,2.5,3.0$. A number of walkers $w \in [W -\Delta W,W +\Delta W] $, with $W$ always chosen equal to half the largest connected component size, is initially randomly placed on the network. Walkers perform 100 walks of $M=1000$ steps each. We monitor the number of visits on each node at the end of the $M$ steps obtaining a series of fluxes for each node. We calculate the mean flux $\langle f \rangle$ and the flux standard deviation $\sigma$ for each node.

 \begin{figure}
        \begin{center}
           \includegraphics[angle=0,width=9.5cm]{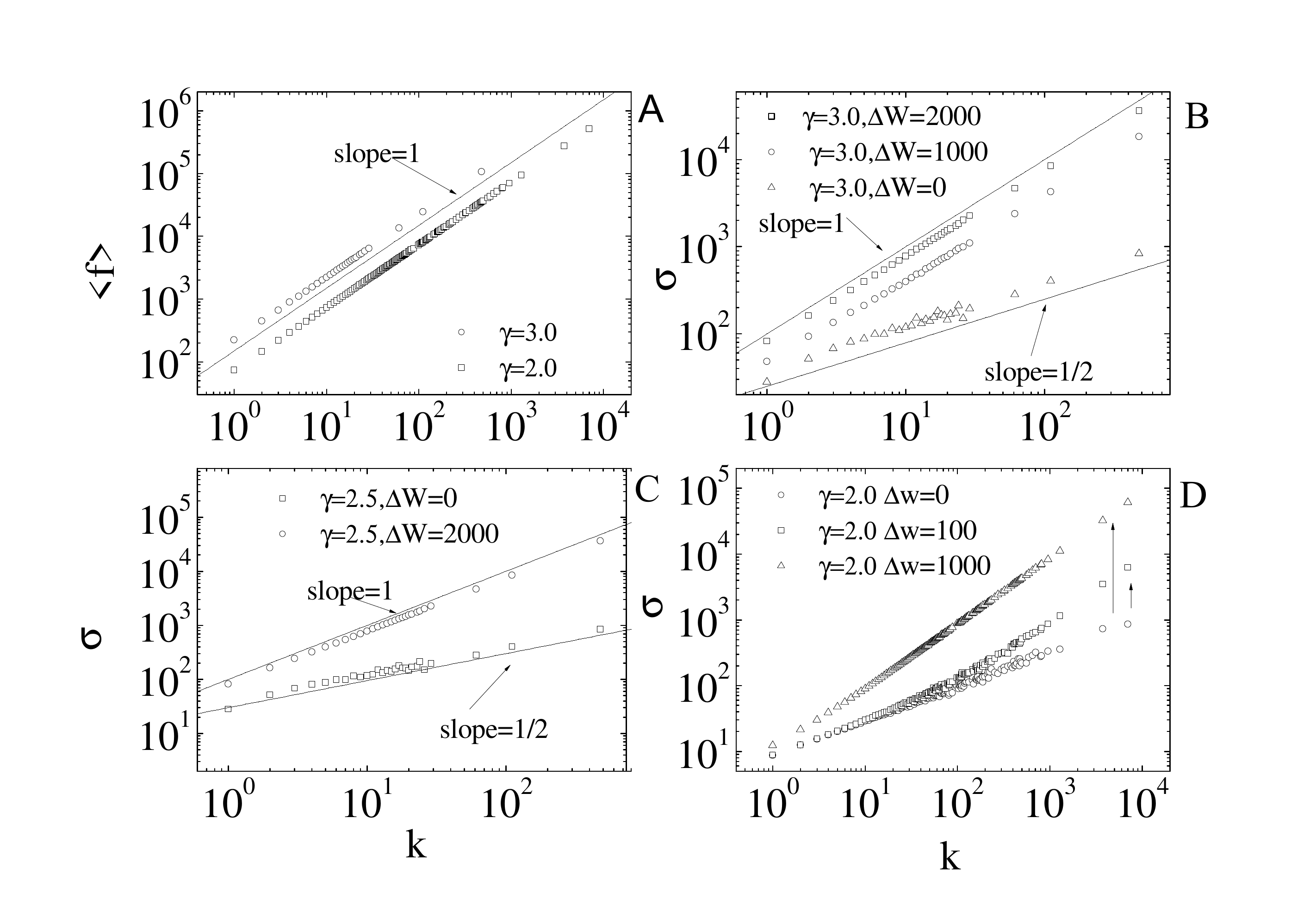}
        \end{center}
        \caption{Dispersion and flux versus degree.(A) Mean flux $\langle f \rangle$ as a function of the node degree $k$ for networks with $N=10^4$, $\gamma=2.0,3.0$ and $\Delta W=0$.(B)Flux standard deviation $\sigma$ as a function of the node degree $k$ for networks with $\gamma=3.0$ and $\Delta W=0,1000,2000$ (C) Flux standard deviation $\sigma$ as a function of the node degree $k$ for networks with $\gamma=2.5$ and $\Delta W=0,2000$.(D) Flux standard deviation $\sigma$ as a function of the node degree $k$ for networks with $\gamma=2.0$ and $\Delta W=0,100,2000$.   }
        \label{fig2}
       \end{figure}
       
Fig.\ref{fig2}A shows the mean flux $\langle f \rangle$ as a function of the node degree $k$. As intuitively expected $\langle f \rangle$ is proportional to $k$ reflecting the fact that nodes with higher connectivity are more frequently visited. This result is valid for all scale-free exponents $\gamma$, thus, we observe straight lines with slope equal to 1 in this double logarithmic plot. It is also in agreement with a theoretical derivation presented in \cite{noh2004random}. Figs.\ref{fig2}B and C show the flux standard deviation $\sigma$ as a function of the node degree $k$ for networks with $\gamma=3.0$ and $\Delta W=0,1000,2000$(top,right) and  $\gamma=2.5$ and $\Delta W=0,2000$. We observe that $\sigma $ is well described as a power law of the degree $k$ of the node. The observed slopes are 1/2 for $\Delta W=0$ and 1 for $\Delta W=1000,2000$. This is in accordance with the intuition gained from the exact results obtained for the star network and the fact that $\langle f \rangle$ is proportional to $k$. Fig.\ref{fig2}D shows the same for $\gamma=2.0$ and $\Delta W=0,100,2000$. In this case, due to the broad $k$ distribution one can see the different sensitivity of the nodes to external noise. The middle curve shows one regime of high-degree nodes that have been considerably affected by $\Delta W$ and another regime of low-degree nodes that remain practically unaffected with a data collapse of the low part of the curves for $\Delta W=0,100$. Arrows indicate the considerable `shift' of $\sigma$ of the large degree nodes when  $\Delta W$ is increased.  

 \begin{figure}
  	 \begin{center}
        	 \includegraphics[angle=0,width=9cm]{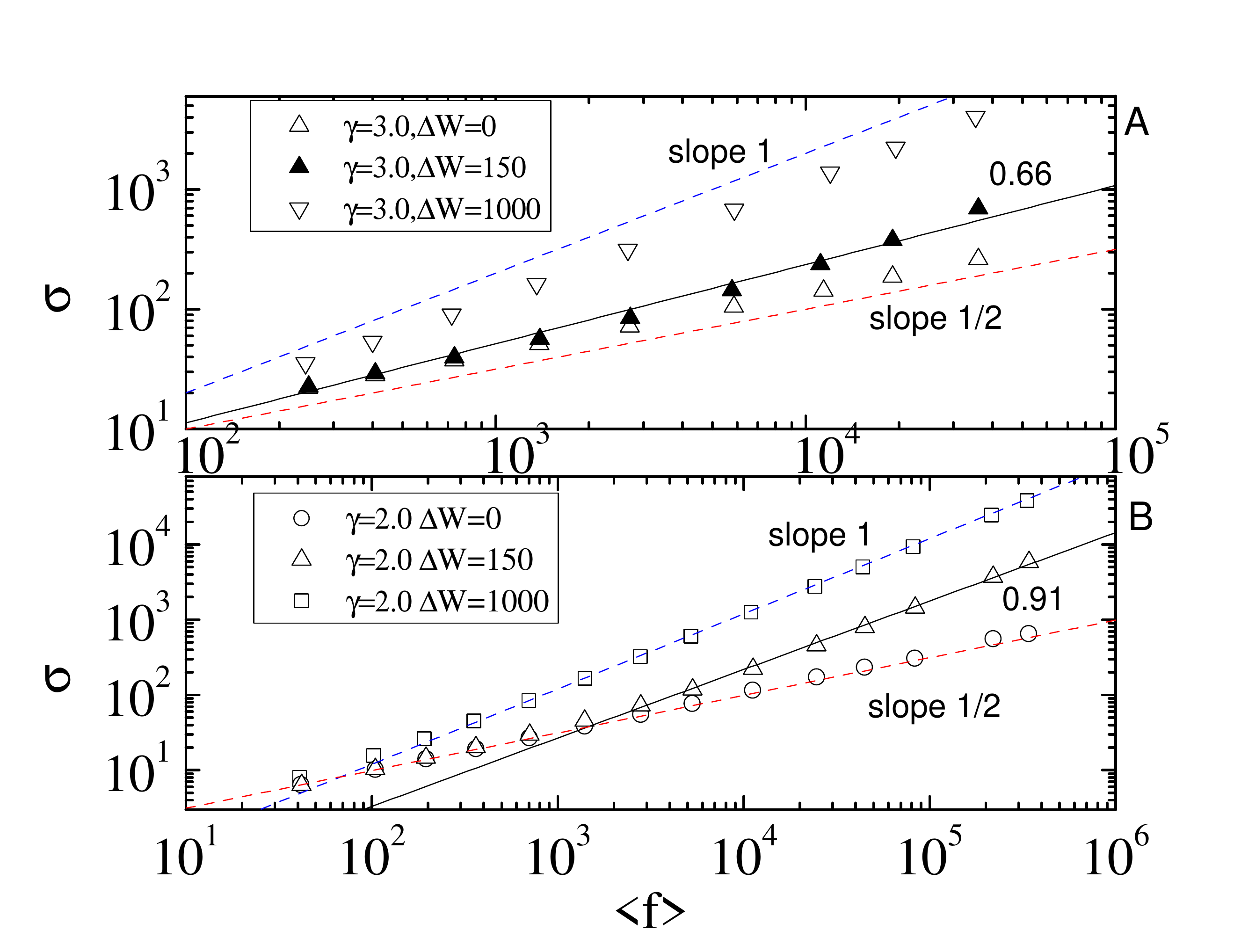}
     \end{center}
         \caption{(A) Node flux standard deviation $\sigma$ as a function of the mean flux $\langle f \rangle$ for scale free networks with $N=10^4$, $\gamma=3.0$ and $\Delta W=0,150,1000$. (B) Node flux standard deviation $\sigma$ as a function of the mean flux $\langle f \rangle$ for scale free networks with $N=10^4$, $\gamma=2.0$ and $\Delta W=0,150,1000$ }
         \label{fig3}
 \end{figure} 

In Fig.\ref{fig3} we plot the node flux standard deviation $\sigma$ as a function of the mean flux $\langle f \rangle$ for scale free networks with $N=10^4$ and $\gamma=3.00$ (Fig.\ref{fig3}A), $\gamma=2.0$ (Fig.\ref{fig3}B) for different values of $\Delta W$. We observe an intermediate regime of curves with slopes considerably different from 1/2 or 1 depending on the value of $\Delta W$. 
In the case of $\gamma=3.0$ one may consider a single power law with slope 0.66 that describes adequately the simulation data of $\Delta W=150$ for the whole range of $\langle f \rangle$ with the exception of the very low fluxes. For $\gamma=2.0$ and $\Delta W=150$ there is obviously a cross-over from low $\langle f \rangle$  data that scale with an exponent 1/2 to a regime of high $\langle f \rangle$ that scale with an exponent $\simeq 0.91$. 

Thus, it is obvious that a single power law is not an adequate description of the flux-dispersion relation in all cases. It may be sufficient, however, when someone is interested in the behavior at the asymptotic limit of large fluxes.
While an exact result is not available for large scale free networks, in contrast to the star network case, Eq.\ref{eq1b} is a rather plausible alternative as shown in \cite{meloni2008scaling}. There, the authors approximate the arrivals on a node when $\Delta W=0$ with a Poisson process. Then the case of $\Delta W \ne 0$ is treated as a mixture distribution and Eq.\ref{eq1b} for the variance as a function of the flux is derived.
Eq.\ref{eq1b} is expected to hold when the arrival statistics for $\Delta W=0$ are not considerably different from that of a Poisson distribution i.e. for large networks with an adequate number of steps performed by the walkers.
The theoretically predicted\cite{meloni2008scaling} value of the parameter $c$ is $(\Delta W/W)^2$.

\begin{figure}
  	 \begin{center}
        	 \includegraphics[angle=0,width=9cm]{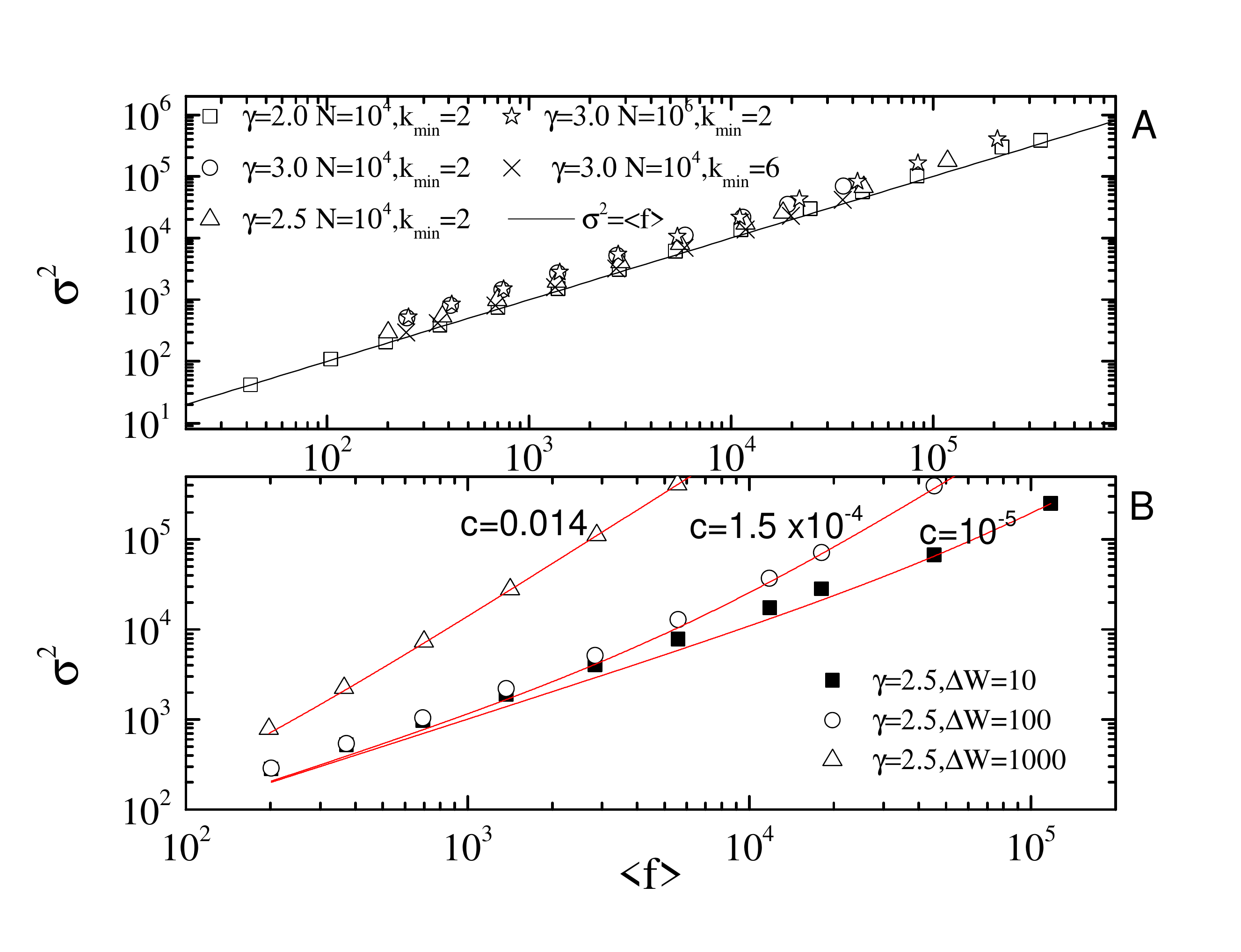}
     \end{center}
         \caption{(A) Variance $\sigma^2$ versus flux $\langle f \rangle$ for
         networks with $\gamma=2.0,2.5,3.0, k_{min}=2$ and size $N=10^4$ (squares,circles,triangles),  $\gamma=3.0, k_{min}=2$ and size $N=10^6$ (stars) and  $\gamma=3.0, k_{min}=6$ and size $N=10^4$ (cross)
         for walks with $\Delta W=0$. The line is the theoretical prediction 
         $\sigma^2=\langle f \rangle $.  
          (B) Variance $\sigma^2$ versus flux $\langle f \rangle$ for
          networks with $\gamma=2.5, k_{min}=2$ and size $N=10^4$ for 3 different $\Delta W =10,100,1000$. Lines are fitting to Eq.\ref{eq1b}   }
         \label{fig4}
 \end{figure} 
 
 In Fig.\ref{fig4} we plot the results of large scale Monte Carlo simulations of random walks on scale-free graphs in order to examine the range of validity of Eq.\ref{eq1b}. Fig.\ref{fig4}A shows  $\sigma^2$ versus flux $\langle f \rangle$ for  networks with $\gamma=2.0,2.5,3.0, k_{min}=2$ and size $N=10^4$ (squares,circles,triangles),  $\gamma=3.0, k_{min}=2$ and size $N=10^6$ (stars) and  $\gamma=3.0, k_{min}=6$ and size $N=10^4$ (cross) for walks with $\Delta W=0$.
  The line is the theoretical prediction $\sigma^2=\langle f \rangle $ i.e. Eq.\ref{eq1b} for $\Delta W =0$ ($c=0$) according to \cite{meloni2008scaling}. This equality is a direct consequence of the well known property of the Poisson distribution to have a variance equal to its mean value. 
  
  Note that the plot scale is doubly logarithmic. Although the vertical distance of the points from the line may seem small on visual inspection it is actually quite significant due to the log scale of the y-axis. 
  
   We see that for networks with $\gamma=2.0, k_{min}=2$ and $\gamma=3.0, k_{min}=6$ the points fall on the straight line indicating that $\sigma^2=\langle f \rangle $ is valid. For $\gamma=2.5,3.0, k_{min}=2$, however, the vertical distance from the line is quite different from zero even for very large networks with $N=10^6$ nodes and one may observe $\sigma^2 \simeq 3\langle f \rangle $. The reason for this difference is rooted in the discrete nature of the arrival statistics which as can be seen from the star network is actually described by a Binomial distribution. The Binomial distribution is known to coincide to a Poisson distribution when the probability of success (in this case arrival) tends to zero. For networks with large average degree (i.e. $\gamma=2.0, k_{min}=2$ and $\gamma=3.0, k_{min}=6$) the mean arrival probability is small and we see a good agreement with the theoretical derivation which is based on the assumption of a Poisson distribution of the arrivals.
  In Fig. \ref{fig4}B we plot $\sigma^2$ versus flux $\langle f \rangle$ for networks with $\gamma=2.5, k_{min}=2$ and size $N=10^4$ for 3 different $\Delta W =10,100,1000$. Lines are fitting to Eq.\ref{eq1b} with one adjustable parameter, the parameter $c$. We see that Eq.\ref{eq1b} is a very effective way of analyzing the data compared to a power law fitting (Eq. \ref{eq1}). In any case, we believe that using a power law in order to describe such data, as is routinely done in the literature is legitimate providing one keeps in mind that it is an approximate law and is usually a sufficient description only if one is interested mainly for the behavior when $\langle f \rangle $ is large.

  \begin{figure}
    \begin{center}
    	\includegraphics[angle=0,width=9.5cm]{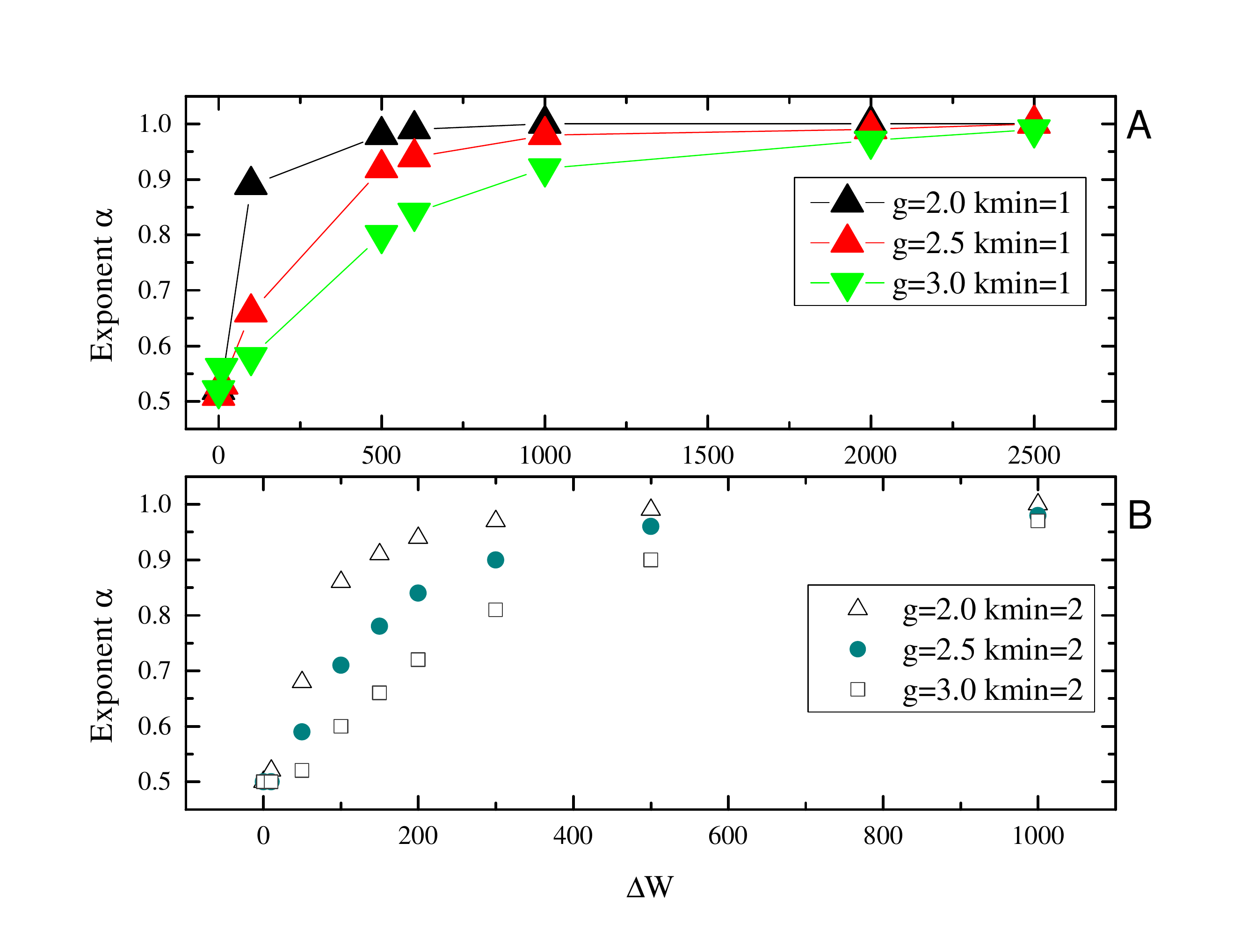}
    \end{center}
    \caption{(A) Exponent $\alpha$ of the dispersion-flux relation as a function of  $\Delta W$ for networks with $\gamma=2.0,2.5,3.0$ and minimum degree $k_{min}=1$. (B) The same but for networks with minimum degree $k_{min}=2$}
    \label{fig5}
   \end{figure}

Using such a power law may indeed elucidate some characteristic properties of the system. Fig.\ref{fig5} shows Monte Carlo simulation results for the exponent $\alpha$ of the dispersion-flux relation (Eq.\ref{eq1}) as a function of the ``external'' noise parameter $\Delta W$ for networks with $\gamma=2.0,2.5,3.0$ and minimum degree $k_{min}=1$ (Fig.\ref{fig5}A) or  $k_{min}=2$ (Fig.\ref{fig5}B). The choice $k_{min}=2$ makes the largest cluster of the networks equal to the network size $N$. Thus, the mean number of walkers is identical no matter the $\gamma$ exponent. Also in the case of $k_{min}=1$ we have used different network sizes so that their largest connected components $N_{lc}$ have roughly equal sizes. In the simulated cases presented in Fig.\ref{fig5}A we had $N_{lc}\simeq 9700$ for $\gamma=2.0$, $N_{lc}\simeq 9500$ for $\gamma=2.5$ and  $N_{lc}\simeq 10300$ for $\gamma=3.0$. We observe that the exponent $\alpha$ approaches 1 with different rates for different $\gamma$ exponents and that the more heterogeneous networks, i.e. those with lower $\gamma$ exponents amplify the effect of the noise parameter $\Delta W$. This is in accordance to our expectations from the solution on the star network, since scale-free networks with $\gamma=2.0$ may be thought of as (connected) collections of many `stars' while for $\gamma=3.0$ those stars are fewer and smaller.

\begin{figure*}
  	 \begin{center}
        	 \includegraphics[angle=0,width=13cm]{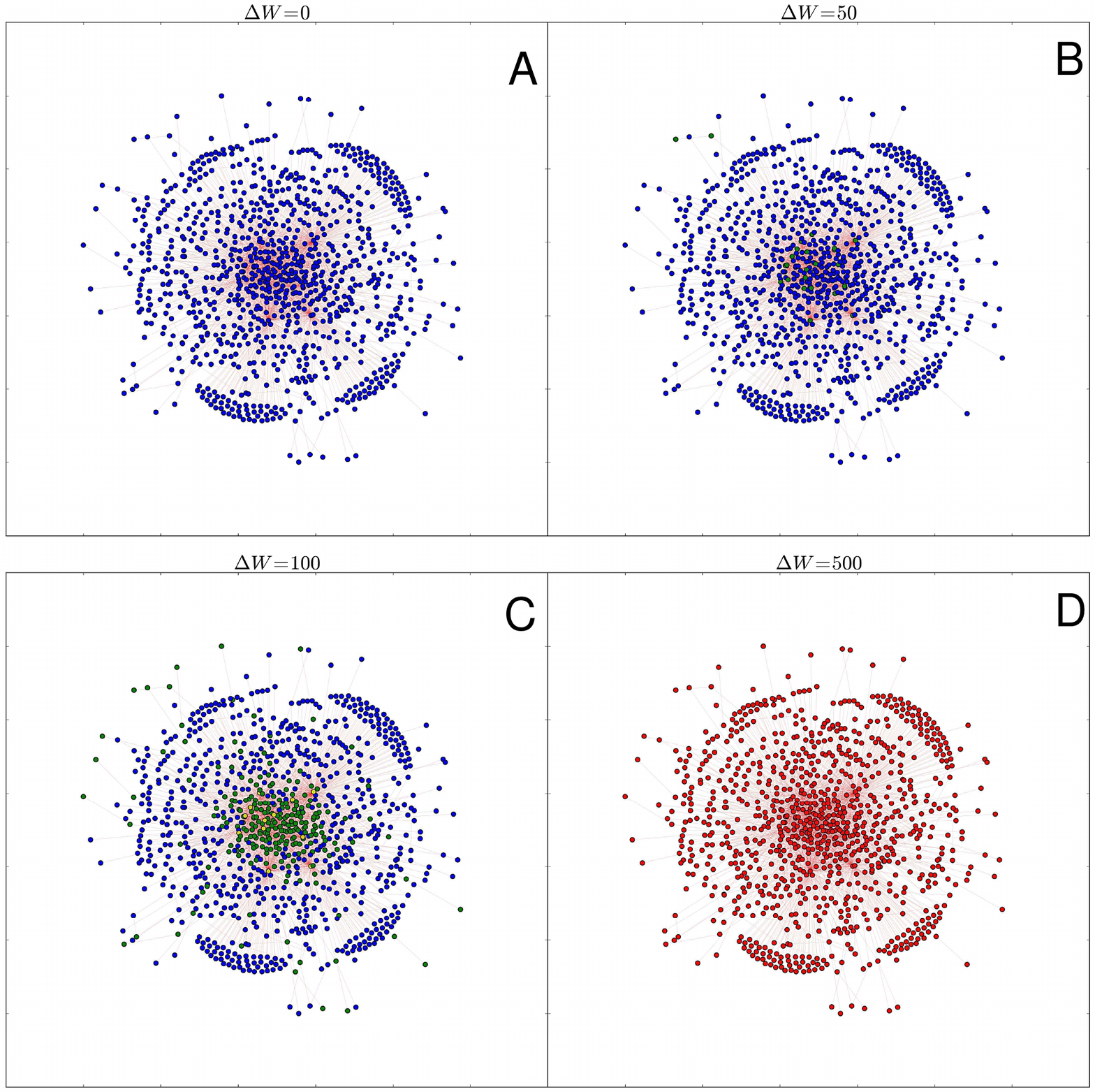}
     \end{center}
         \caption{Scale-free network with 1000 nodes $\gamma=2,k_{min}=1$ for $\Delta W=0,50,100,500$. Edges are depict as red lines connecting nodes. The color of the nodes depends on the value $r_i=\frac{\log \sigma}{\log f}$ of each node. Nodes with $r_i<0.6$ are blue, $0.6<r_i<0.75$ are green,$0.75<r<0.85$ are yellow and $r_i>0.85$ are red. }
         \label{fig6}
 \end{figure*} 
 
The parametrization Eq.\ref{eq1} is rather useful for visualization purposes as it allows for a parameter $r_i=\frac{\log \sigma_i}{\log f_i}$ to be associated with each node. In Fig.\ref{fig6} we plot a scale-free network with 1000 nodes $\gamma=2,k_{min}=1$ and $\Delta W=0,50,100,500$. The color of the nodes has been chosen according to the value $r_i=\frac{\log \sigma}{\log f}$ of each node $i$. Note that in contrast to the exponent $\alpha$ of Eq.\ref{eq1} which represents a collective graph property, the quantity $r_i$ signifies a node property, hence the subscript $i$. Actually, $r_i$ can be interpreted as the `effective' scaling of node $i$.   Nodes with $r_i<0.6$ are shown as blue, $0.6<r_i<0.75$ are green,$0.75<r_i<0.85$ are yellow and $r_i>0.85$ are red. A `spring layout' algorithm has been used to position the network nodes and thus nodes with high degree are placed towards the center of the figure. For $\Delta W=0$ is rather homogeneous with all nodes towards the `blue' spectrum. For $\Delta W=50$ we observe several well connected nodes towards the center to appear with yellow color indicating higher $r_i$ values. These are the nodes that have been affected by the introduction of the `external' noise. The effect becomes much more dramatic for $\Delta W=100$ with most of the central nodes having higher $r_i$ values and a characteristic pattern of yellow central nodes appearing. For $\Delta W=500$ all nodes are actually affected from external noise, the figure is rather homogeneous again but all nodes are towards the `red' spectrum values $r_i \simeq 1.0$.          

\begin{figure}
  	 \begin{center}
        	 \includegraphics[angle=0,width=9cm]{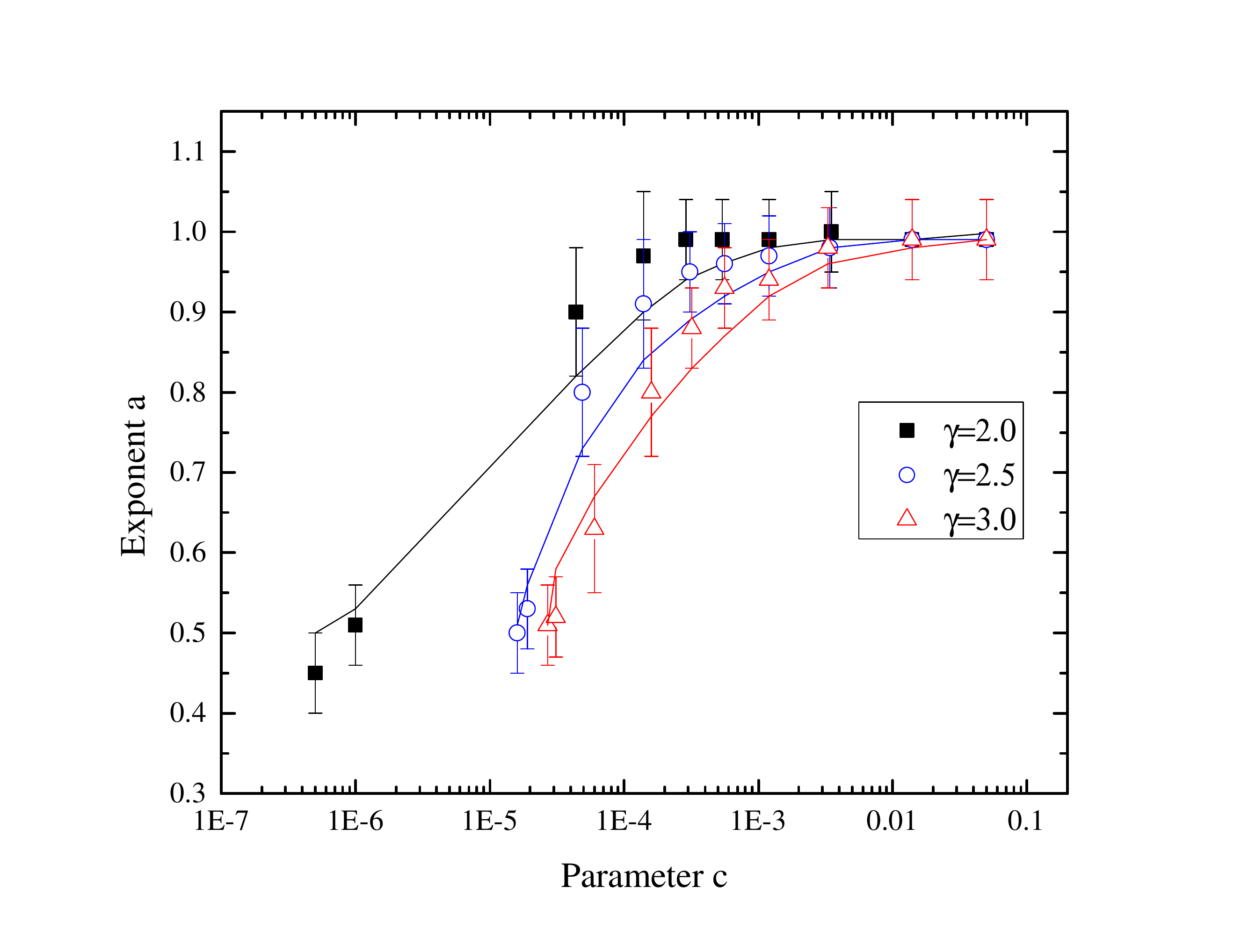}
     \end{center}
         \caption{Exponent $a$ versus parameter $c$ for networks with $\gamma=2.0,2.5,3.0, k_{min}=2$ and size $N=10^4$ (squares,circles,triangles). Lines are predictions of Eq.\ref{eq15}  }
         \label{fig7}
 \end{figure} 

Since, moreover, a power law is often used to analyze flux-dispersion data in the literature it would be instructive, and potentially useful in practice, to see how the parameter $c$ of Eq.\ref{eq1b}
and the exponent $a$ of Eq.\ref{eq1} are related. We can derive an equation in closed form as follows:
Let $y_1=\ln{(f+c f^2)}$ and $y_2=\ln{(b f^n)}$. Then we want to estimate the values of $b$ and, most importantly, $a$ that minimize the integral of the squared differences $I=\int_{0}^{m} (y_2-y_1)^2 df $ where $m$ is the maximum flux. The logarithms of $y_1,y_2$ are taken for convenience, otherwise a closed form relation cannot be obtained.
The minimization condition is, of course, equivalent to setting $\partial I/\partial b =0$ and $\partial I/ \partial n =0$.

When $W$ walkers are performing random walks on a network of $N$ nodes , at the stationary regime the average number of walkers on a node with degree $k$ is $k W/(N \bar{k} )$ where $\bar{k}$ is the mean degree of the nodes. We can use this to estimate the maximum flux $m$ as $m=k_{max} s W/(N \bar{k})$ where $s$ is the number of steps that the walkers perform.
We find that the exponent $\alpha=n/2$ of Eq.\ref{eq1} (the factor of 2 comes from the fact that Eq.\ref{eq1} relates the standard deviation to the flux while Eq.\ref{eq1b} the variance to the flux) is related to the parameter $c$ as :

\begin{equation}
\begin{split}
  \alpha= 
  & \frac{1}{12 c m} (6 \text{Li}_2\left(\frac{1}{c m+1}\right)+ 
   12 c m+ \\ 
  & 3 \ln (c m+1) (\ln (c m+1)
   -2 (\ln (c)+\ln (m)))-\pi ^2 )
\end{split}
    \label{eq15}
\end{equation}

  which along with $m=k_{max} s W/(N \bar{k})$ is the desired result connecting the 2 different model parameters to the network topology.
  In Eq.\ref{eq15} $\text{Li}_2$  is the polylogarithmic function defined by  $\text{Li}_n(z)=\sum _{k=1}^{\infty } \frac{z^k}{k^n}$ for $n=2$.
  Figure \ref{fig7} is a plot of the exponent $a$ versus parameter $c$ for networks with $\gamma=2.0,2.5,3.0, k_{min}=2$ and size $N=10^4$ (squares,circles,triangles). Points are estimates of the parameters from fitting of Eq.\ref{eq1} (exponent $\alpha$) and Eq. \ref{eq1b} (parameter c) to Monte Carlo Simulation data. Lines are the predictions of Eq.\ref{eq15}. We observe a rather good agreement between the two showing that, despite it's rather complicated functional form, Eq. \ref{eq15} is a useful tool in connecting the two ways of data analysis.

\begin{figure}
    \begin{center}
      	  \includegraphics[angle=0,width=9.5cm]{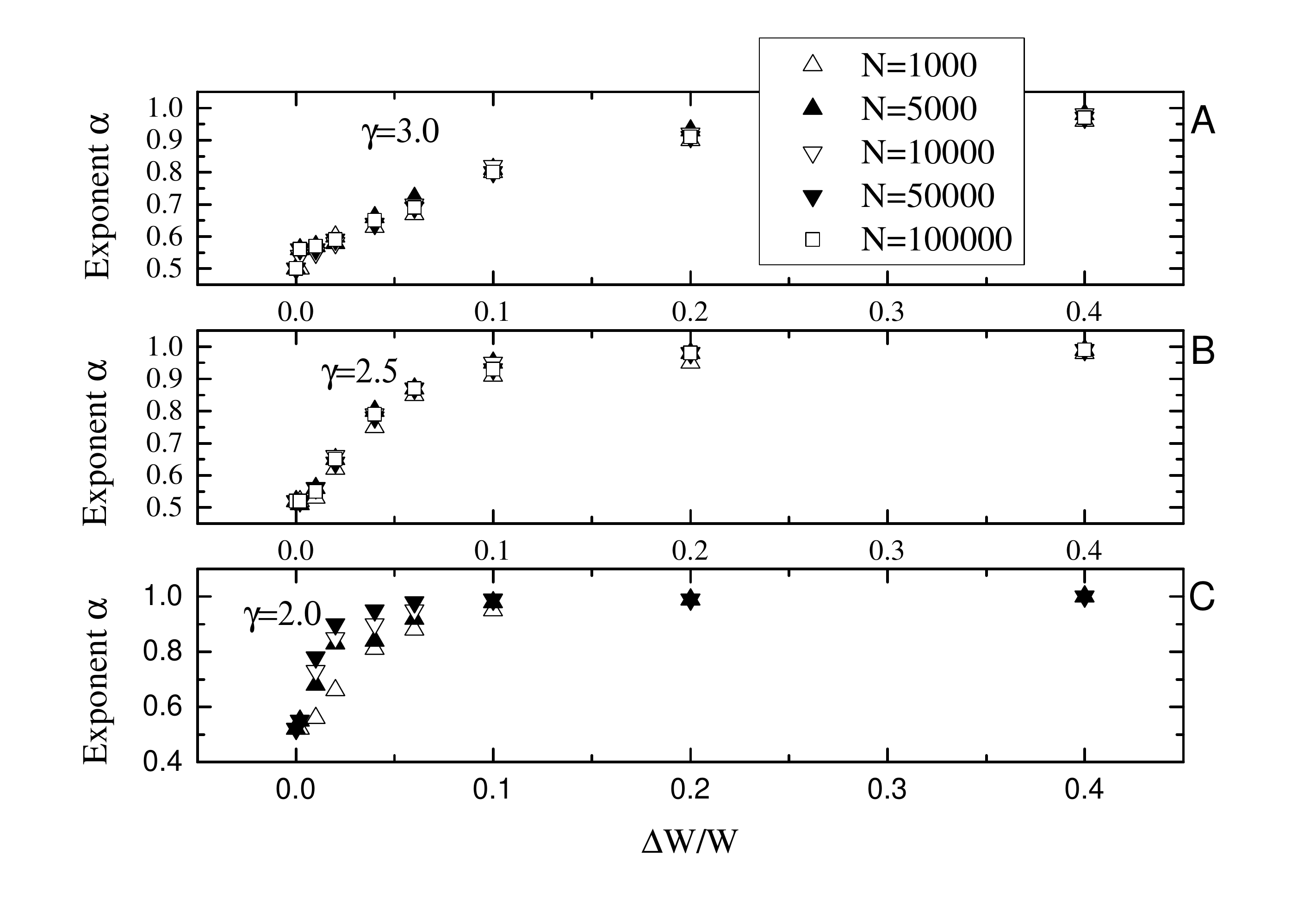}
    \end{center}
    \caption{(A) Exponent $\alpha$ as a function of $\Delta W/W$ for networks of $\gamma=3.0,k_{min}=2$ for different network sizes, namely $N=1000,5000,10000,50000,100000$. (B)The same for networks with $\gamma=2.5,k_{min}=2$ (C) The same for networks with $\gamma=2.0,k_{min}=1$ }
    \label{fig8}
\end{figure}

Equation \ref{eq15} can be used in order to estimate the effect of network size $N$ on the value of the exponent $a$. The parameter $c$ does not depend on the network size. Its theoretical value equals the ratio $(\Delta W/W)^2$ and is, thus, independent of $N$ and  our simulation results confirm this independence within statistical errors. The exponent $\alpha$ is, however, dependent on $N$ since it is a function of the maximum node degree $k_{max}$ through the dependence of $m$ to  $k_{max}$ ,(see Eq.\ref{eq15}). Note that, this effect may be difficult to observe in moderate network sizes since, for example, a 5-fold increase of $N$ for a network of $\gamma=3.0$ will only lead to a 2-fold (or less) increase of $k_{max}$.   

Fig.\ref{fig8} shows the exponent $\alpha$ as a function of $\Delta W/W$ for networks of $\gamma=3.0,k_{min}=2$ (Fig.\ref{fig8}A), $\gamma=2.5 ,k_{min}=2$ (Fig.\ref{fig8}B) and 
 $\gamma=2.0 ,k_{min}=1$ (Fig.\ref{fig8}C) for different network sizes, namely $N=1000,5000,10000,50000,100000$. We observe that for $\gamma=3.0$ it is not possible to observe an increased exponent with network size in the intermediate regime. For  $\gamma= 2.5$  the data collapse is not so strong and such an increase is observed but still statistical errors do not allow the extraction of clear results just from the simulation data. Eq.\ref{eq15} is a valuable tool in this case showing that the expected increases of the exponent are of the order of $0.05$ which is well within the statistical error for $\alpha$.
 For  $\gamma= 2.0$ one may observe an increase of the exponent in the intermediate regime in agreement with the effect predicted by Eq.\ref{eq15}, as these networks have rather large $k_{max}$ values even for medium sized networks. This result confirms that the intermediate exponent region will not vanish for medium size networks and, thus, signifies an important regime, potentially observable in the analysis of real systems. 
   
\begin{figure}
    \begin{center}
      	\includegraphics[angle=0,width=9.5cm]{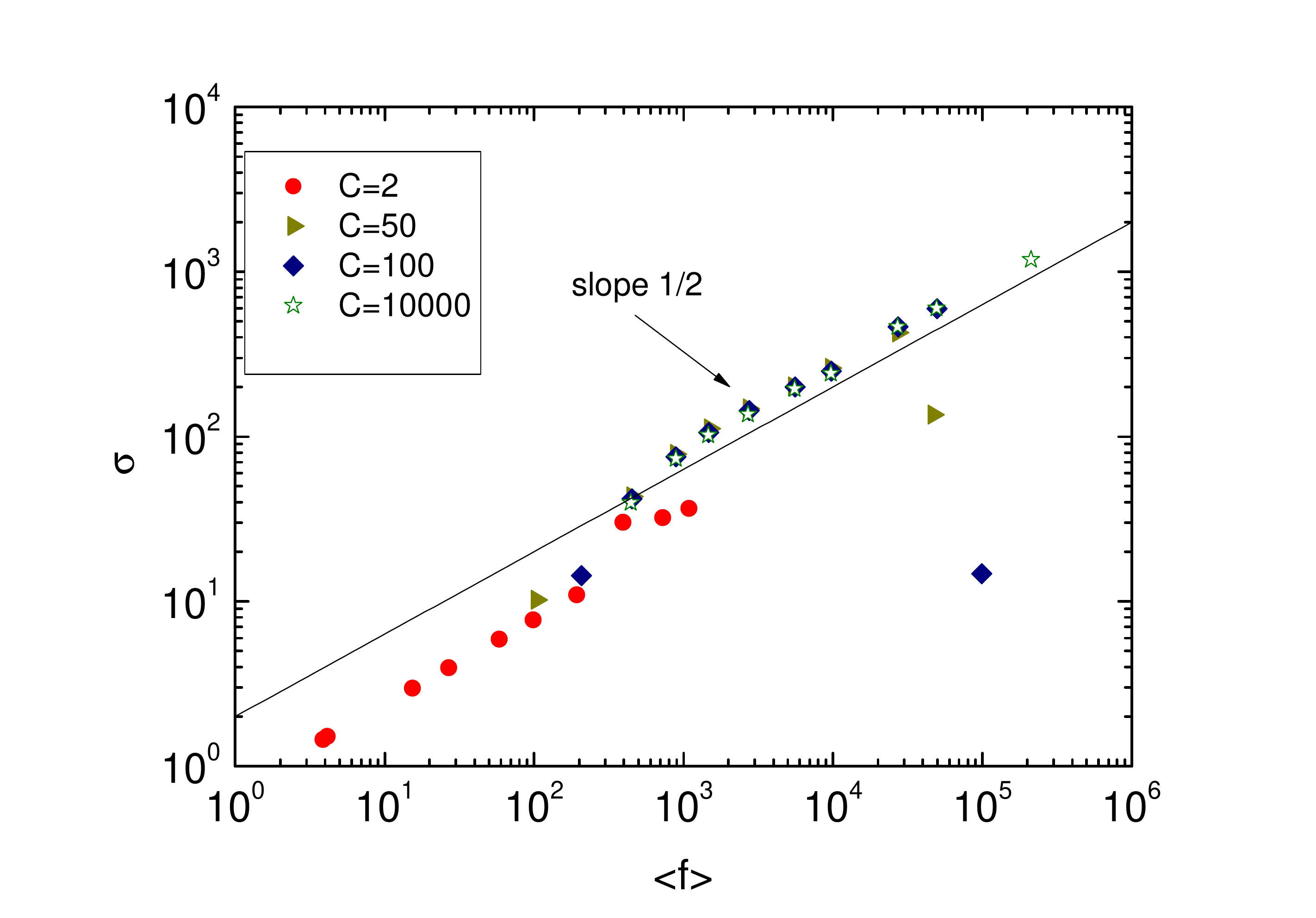}
    \end{center}
    \caption{Flux standard deviation $\sigma$ versus $\langle f \rangle$ for a network with $\gamma=3.0$ with $N=10^4,M=1000$. Largest cluster ($N_{lc} \simeq 7000$) is used. Walkers are initially placed randomly on the network as long as the capacity of a node permits it. } 
    \label{fig9}
\end{figure}

The parameterization Eq.\ref{eq1} assumes a homogeneous network, where all nodes obey the same scaling relationship. The parameterization Eq.\ref{eq1b} is capable of accounting for two groups of nodes with different behaviors. As the average flux $\langle f \rangle$ is proportional to the degree $k$ the two regimes visible in Figs.{\ref{fig2},\ref{fig4}} show that fluctuations at high-degree nodes and at low-degree nodes scale differently.
Can we have more groups of nodes with distinct behaviors under random walks in the network? Surprisingly, this is true if we consider the concept of node capacity. In all the above cases we have assumed that the nodes can support an unlimited number of walkers without performance degradation. In most real-life large scale transport systems, however, the nodes have a limited capacity. Computer routers for example have upper limits on the amount of incoming and outgoing traffic rate that they will support. 
Thus, we are interested in studying the influence of the capacity $C$ on the flux-dissipation relation. For this study we again use scale free network topologies and we additionally assume that each node has a capacity $C$ defined as the maximum number of random walkers that can occupy the node at the same time. When $C=1$ we have the well-known case of excluded volume interactions. For the simulations we have used the largest cluster of a 10000-node scale free network with $\gamma=3.0$. The size of the largest cluster is $N_{lc}\simeq 7000$. The number of walkers placed on the network was set equal to $N_{lc}$ and $\Delta W=0$. The maximum degree $k_{max}$ of the network is $k_{max} \simeq 150$.
Walkers (try to) perform $M=1000$ steps each. At the end of the steps fluxes are recorded for each node. The process repeats for 100 times and time-average fluxes and standard deviations are calculated for each node. In Fig.\ref{fig9}
we plot the flux standard deviation $\sigma$ versus $\langle f \rangle$ for a network with $\gamma=3.0$ with $N=10^4,M=1000$. Walkers are initially placed randomly on the network as long as the capacity of a node permits it. 

We find 3 interesting regimes: 

(a) For $C \simeq (W/N) k_{min} $ the observed flux range is decreased but the power law scaling with exponent 1/2 remains. In this case,(see $C=2$ in Fig.\ref{fig9}) all nodes feel the limitations of the capacity $C$. Walkers will often try to move but will find the destination to be fully occupied. In such a case they will not perform a step, resulting in a reduced number of arrivals on all counters. 

(b) For $C \simeq (W/N) k_{max}$ we notice the appearance of outliers at the beginning and end of the series (see $C=50,100$ in Fig.\ref{fig9}) which do not allow to claim that the scaling relation is well approximated by a power-law. Here the high degree nodes (hubs) are influenced from the capacity limitation. Thus, while the low-degree nodes
receive more or less the same amount of walker arrivals as in the unlimited case
the hubs reach the capacity limit. This limitation significantly alters the visitation pattern and leads to the observed appearance of the outliers. In several cases the walkers will be unable to visit the saturated hubs and thus will remain unmovable for some steps leading to decreased ``flux'' recorded at the counters compared to the unsaturated case.

(c) For $C \gg (W/N) k_{max}$ the power law scaling with exponent 1/2 is, as expected, recovered since the nodes can accommodate all possible arrivals without problem similar to the case of unlimited capacity.

\begin{figure}
    \begin{center}
      	\includegraphics[angle=0,width=9.5cm]{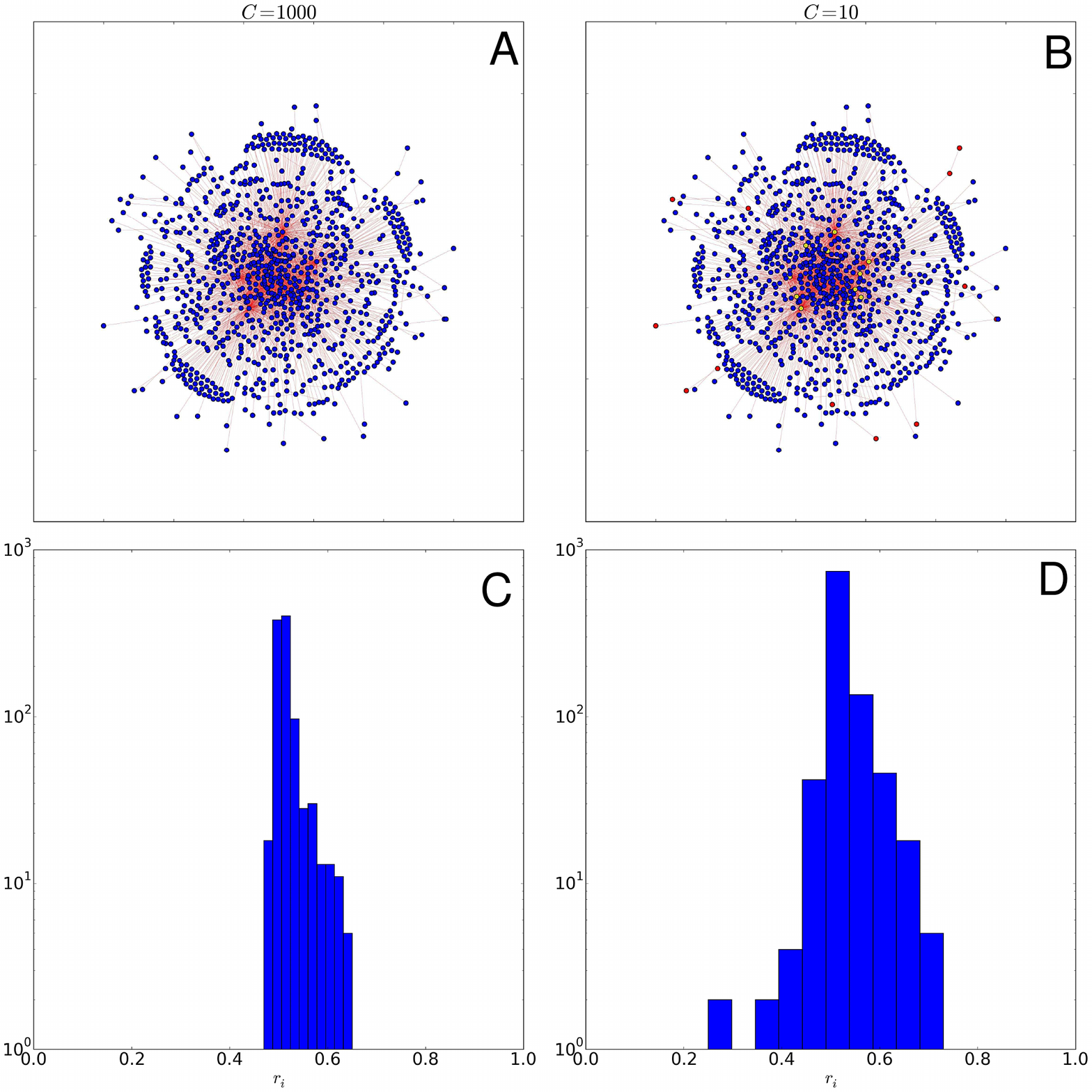}
    \end{center}
    \caption{Scale-free network with 1000 nodes $\gamma=2,k_{min}=1$ for $\Delta W=0$ and capacity $C=1000$(A), $C=10$(B) . Edges are depict as red lines connecting nodes. The color of the nodes depends on the value $r_i=\frac{\log \sigma}{\log f}$ of each node. Nodes with $r_i<0.45$ are yellow, $0.45<r_i<0.65$ are blue,and those with $r_i>0.65$ are red.(C) Histogram of the $r_i$ values for $C=1000$.(D) Histogram of the $r_i$ values for $C=10$.} 
    \label{fig10}
\end{figure} 

Thus, when the capacity parameter is taken into account, we can roughly distinguish 3 types of nodes i.e. high degree saturated nodes, high-medium degree unsaturated nodes and low degree nodes, which influence the fluctuations scaling in quantitatively different ways. In Figure \ref{fig10} we plot 2 extreme capacity cases of a scale-free network  with 1000 nodes $\gamma=2,k_{min}=1$ for $\Delta W=0$ and capacity $C=1000$(Figs \ref{fig10}A,C), $C=10$(Figs \ref{fig10}B,D)) . High degree nodes are placed near the center of the graph. The color of the nodes depends on the value $r_i=\frac{\log \sigma}{\log f}$ of each node $i$. Nodes with $r_i<0.45$ are yellow, $0.45<r_i<0.65$ are blue,and those with $r_i>0.65$ are red. The case $C=1000$ (Fig \ref{fig10}A) is actually a network with internal fluctuations only, since  $\Delta W=0$, and practically unlimited capacity. As expected, the fraction $r_i$ is close to 0.5 for all nodes and all nodes appear with blue color. When $C=10$ (Fig \ref{fig10}B), although $\Delta W=0$ and no external fluctuations are present, we observe the appearance of highly connected saturated nodes (yellow) with $r_i<0.45$ coexisting with unsaturated nodes (blue nodes towards the center of the graph) as well as some nodes with high $r_i$ ratio consistent with our previous remarks. The bottom figures are histograms of the $r_i$ values for the two cases $C=1000$(Fig \ref{fig10}C) and $C=10$ (Fig \ref{fig10}D). Notice the appearance of additional `bands' close to $0.25$ and $0.75$ which are not present in the case of `unlimited' capacity.    

\begin{figure}
    \begin{center}
      	\includegraphics[angle=0,width=9cm]{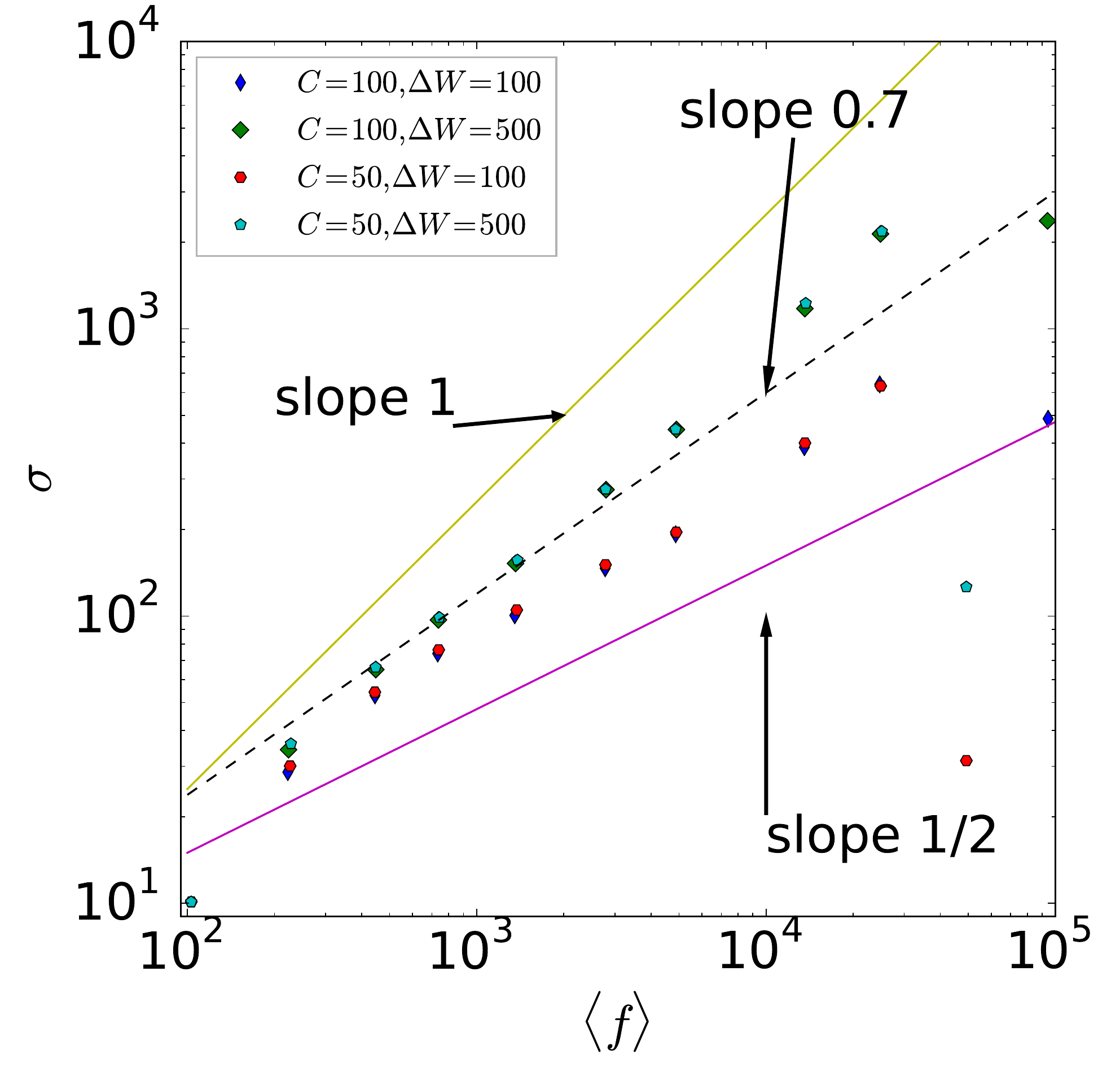}
    \end{center}
    \caption{Flux standard deviation $\sigma$ versus $\langle f \rangle$ for a network with $\gamma=3.0$ with $N=10^4,M=1000$. Different capacity $C$ and $\Delta W$ combinations are shown.} 
    \label{fig11}
\end{figure}  

In case we include some external noise $\Delta W$ in addition to the capacity
we expect to observe a combination of slope changing due to external noise and the appearance of `outliers' i.e. unconventional points of the $\sigma-f$ plot due to the saturated nodes. In Figure \ref{fig11} we plot the  $\sigma$ versus $\langle f \rangle$ for a network with $\gamma=3.0$ with $N=10^4,M=1000$ for different capacity $C$ and $\Delta W$ combinations, namely $C=100,\Delta W=100$(small diamonds), $C=100,\Delta W=500$(large diamonds), $C=50,\Delta W=100$(circles), $C=100,\Delta W=500$(stars).
We can again observe the influence of the external noise which is increasing the
slope of the curves with increasing $\Delta W$. Especially for the cases of $\Delta W=500$ we observe that for $\langle f \rangle < 10^3$ there is a regime with slope around 0.7 (dashed line) due to the influence of the external noise on the low degree nodes, a second regime with a higher slope due to the influence of the external noise on the high degree unsaturated nodes and out-liner points due to the high degree saturated nodes.  

\section{Conclusions}
Stochastic processes on networks exhibit fluctuations due to combinations of internal and external noise. 
We have used a multiple random walk model to study the effect of network heterogeneity on the fluctuations of network dynamics and used random walks as a tool for understanding the relationship between topology and dynamics. We have obtained exact results for the star network which are indicative of the behavior of large scale-free networks. We have found that the network heterogeneity acts as an amplifier of the effects of external noise. These effects include a range of exponents between $1/2$ and $1$ and are persistent for medium size networks. Moreover, we propose that an analysis of the flux variance as a sum of 2 power laws .i.e. a relation of the form $\sigma^2=\langle f \rangle +c (\langle f \rangle)^2 $ with only one adjustable parameter $c$ can provide a more adequate description of the problem under investigation. In particular, we derive a semi-analytical expression for the error of `data' following equation \ref{eq1b}, when describing them with Eq.\ref{eq1}. In this way we can understand (i) why the `transition' from $\alpha=1/2$ to $\alpha=1$ at increasing $\Delta W$ is \textit{not} becoming sharper with increasing network size, and (ii) why network heterogeneity amplifies the influence of external noise. Finally, we have shown that when the internal dynamics correlate with the external influence, as in the case of nodes with a maximum capacity, there appear regimes with non-power law scaling characterized by the appearance of outliers.

An alternative implementation of a finite capacity is to allow queues to form at nodes in times, where the total capacity of the node has been exceeded. Such an extension of random walks to queuing has been discussed in \cite{duch2006scaling}. 

In that case, one can also resort to various analytical results in queueing theory. For example, an interesting alternative treatment of the scaling crossover in the case in which network dynamics is best modeled in terms of a network of waiting lines can be found in \cite{jackson1957networks} . There, a system of $M$ departments is considered, each one with it's own mean arrival rate and mean serving time.
Customers from outside the system arrive at a department $k$ in a poisson type manner and when served are moved (instantaneously) to another department $m$ with a certain probability $\theta(k,m)$ or the service is completed. The steady-state queue length of such a system is shown to follow a negative binomial (Pascal) distribution. Assuming that the flux is proportional to the queue length, the flux-fluctuations relation can be controlled by changing $\theta(k,m)$ from 0 (non-interacting departments) to 1 (fully interacting departments). For non-interacting departments the arrivals are a poisson process, hence leading to exponent 1/2 while the fully interacting case is expected to lead to exponent 1 for appropriate choices of the arrival rates and serving times.   

Exploring scaling relations at the intersection of these two topics, queing theory and random walks on graphs, further can lead to additional interesting insights into the `patterns' on a graph formed by the different scaling behaviors.

\bibliographystyle{apsrev}
\bibliography{flux_v10}

\end{document}